\documentclass{Interspeech}

\makeatletter
\def\maketitle{\par
 \begingroup
    \if@twocolumn
        \twocolumn[\@maketitle] 
    \else%
        \newpage
        \global\@topnum\z@ \@maketitle 
    \fi

    \ifx\@thanks\@empty\else 
       \@thanks 
       \global\let\@thanks\@empty 
    \fi

    \ifinterspeechfinal
        \ifequalcontribution
            \def\thefootnote{\textdagger} 
            \footnotetext{These authors contributed equally.}
            \def\thefootnote{\arabic{footnote}} 
        \fi
    \fi
 \endgroup
}
\makeatother

\interspeechcameraready

\title{SiamCTC: \\ Learning Speech Representations through Monotonic Temporal Alignment }

\author[affiliation={1}]{SooHwan}{Eom}
\author[affiliation={2}]{Mark}{Hasegawa-Johnson}
\author[affiliation={1}]{Chang D. }{Yoo \noexpand\sthanks{Corresponding author.}}

\affiliation{}{Korea Advanced Institute of Science and Technology}{South Korea}
\affiliation{}{University of Illinois Urbana-Champaign}{United States}
\email{\{sean1105, cd\_yoo\}@kaist.ac.kr, jhasegaw@illinois.edu}

\usepackage{comment}
\usepackage{mathtools}
\usepackage{multirow}
\usepackage{xcolor}
\usepackage{booktabs}
\usepackage{pifont}
\newcommand{\cmark}{\ding{51}}  
\newcommand{\xmark}{\ding{55}}  

\begin{document}

\maketitle
\renewcommand{\thefootnote}{\arabic{footnote}}
\setcounter{footnote}{0} 

\begin{abstract}
Self-supervised speech representation learning has made significant progress through Siamese networks, which leverage different views of the same input. However, existing methods often require frame-wise alignment between these views, overlooking the broader linguistic context invariance across different speaking styles. We introduce SiamCTC, a framework that integrates Siamese networks with Connectionist Temporal Classification (CTC) to learn speech representations without strict frame-level correspondence. By employing CTC loss to establish flexible, monotonic alignments between differing temporal realizations of the same content, SiamCTC accommodates speed perturbations and other temporal augmentations. This design relaxes frame-wise constraints while preserving temporal coherence and enhancing robustness to speaking-rate variations in downstream tasks. Our experiments demonstrate that SiamCTC leads to more adaptable speech representations, particularly at diverse speaking rates.
\end{abstract}

\section{Introduction}
\label{sec:intro}
Self-supervised learning (SSL) for speech representation learns from unlabeled data, using the input signal itself as the supervisory signal. By avoiding manual transcriptions, SSL enables deep neural network training on large-scale raw speech corpora, providing effective pre-training and robust representations for downstream tasks such as automatic speech recognition (ASR) \cite{asr} and speaker verification \cite{sv}.

Speech signals inherently capture a variety of attributes, including speaker characteristics, environmental conditions, and linguistic content. Our primary focus is on extracting linguistic context, specifically, latent phonetic properties that characterize linguistic information. Previous SSL approaches have employed techniques such as in-utterance contrastive learning \cite{cpc, wav2vec, wav2vec2} or masked unit prediction \cite{hubert}. However, they often neglect linguistic content invariance to varying speech conditions (e.g., speaking rate), limiting robustness when test conditions differ from those seen during training \cite{robust-wav2vec2}.

Motivated by the growing success of self-supervised learning in computer vision, we turn to Siamese-based frameworks \cite{simclr, dino, byol, moco, simsiam}, which learn from two distinct `views' of the same data sample. These frameworks employ objectives such as contrastive learning, similarity maximization, or masked feature prediction to capture the shared information between views. A key insight of these methods is the existence of a latent structure that remains invariant under different transformations, which Siamese networks aim to extract.

Several attempts have been made to apply Siamese networks to speech representation learning \cite{spin, rspin, csiam, dinosr}. However, they typically rely on frame-wise self-supervised labels, creating two main limitations: (1) restricting temporal augmentations like speed perturbation, as these can introduce frame misalignment, and (2) not fully leveraging linguistic content invariance across natural variations in speaking style and rate. 

To overcome these limitations, we propose \textbf{SiamCTC}, a novel framework for speech representation learning that uses monotonic alignment between latent representations and self-supervised labels. Our approach integrates three key components: (1) a Siamese network to extract invariant speech representations, (2) a Connectionist Temporal Classification (CTC) \cite{ctc} prediction head to learn monotonic alignments without imposing strict frame-wise constraints, and (3) in-utterance temporal contrastive learning to prevent representation collapse.

By explicitly modeling content invariance within utterances, SiamCTC captures linguistic representations more effectively, even with misaligned views introduced by speed perturbations. Our main contributions are summarized as follows:
\begin{itemize}
    \item A flexible framework that eliminates frame-wise alignment constraints while maintaining temporal coherence through CTC-based monotonic alignment learning.
    \item An effective combination of Siamese networks and CTC for robust speech representation learning, improving generalization to varied speaking styles and speeds.
    \item Comprehensive empirical evaluation across multiple downstream tasks, improving the widely used SSL models such as HuBERT \cite{hubert} and WavLM \cite{wavlm}.
\end{itemize}

Through extensive experiments, we demonstrate that SiamCTC achieves improved results compared to conventional SSL frameworks, highlighting the benefits of speech invariance through monotonic alignment.

\begin{figure*}[t!]
    \centering
    \includegraphics[width=0.9\linewidth]{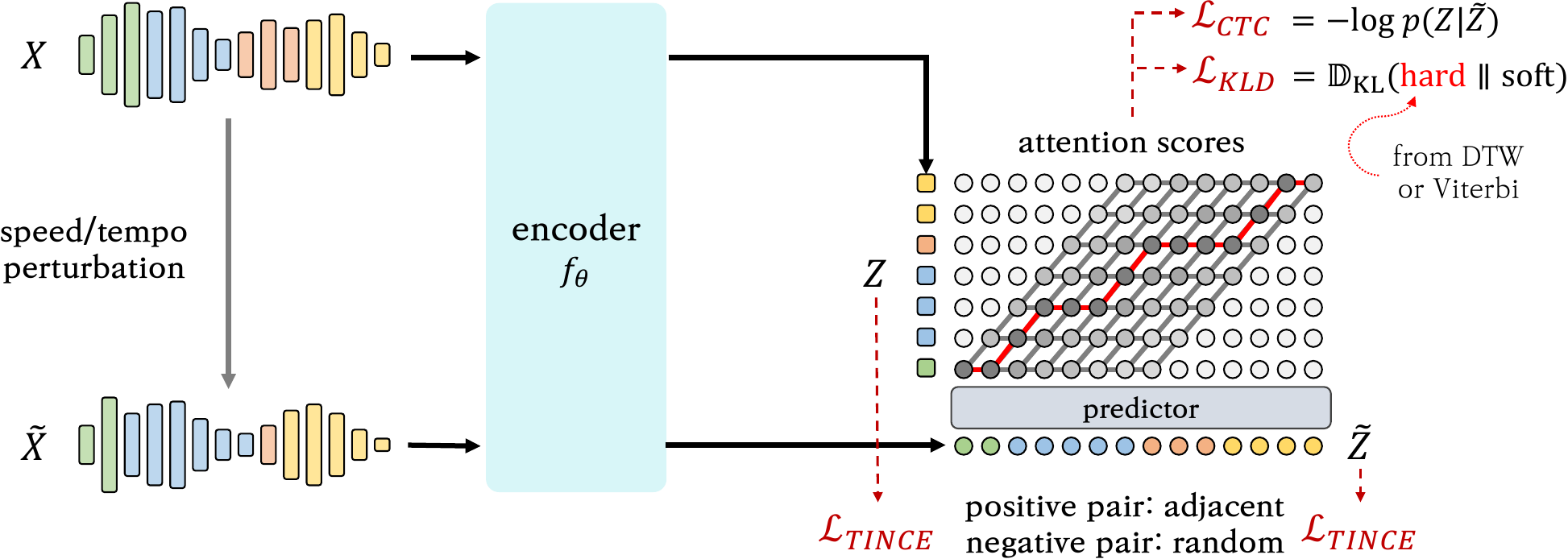}
    \caption{\textbf{An overview of SiamCTC framework.} Overview of our proposed SiamCTC framework. The model processes two views of the same input sequence $X$ and $\tilde{X}$, where one undergoes speed/tempo perturbation. Both sequences pass through a shared encoder $f_\theta$ to produce representation $Z$ and $\tilde{Z}$. The framework optimizes three learning objectives: (1) CTC loss ($\mathcal{L}_{CTC}$) for monotonic alignment learning, (2) KL divergence loss ($\mathcal{L}_{KLD}$) between hard and soft alignments, and (3) Temporal InfoNCE losses ($\mathcal{L}_{TINCE}$) that treat adjacent features as positive pairs while sampling negative pairs randomly within the sequence. Hard alignments are computed via the Viterbi algorithm \cite{viterbi} or DTW \cite{dtw} and serve as a reference for the soft alignments learned by the model.}
    \vspace{-0.5cm}
    \label{fig:siamctc}
\end{figure*}

\section{Related Works}
\label{sec:background}

\subsection{Self-Supervised Speech Representation Learning} Self-supervised speech representation learning methods can be broadly classified into three categories based on their pretext task. Generative approaches (e.g., VQ-VAE \cite{vqvae}, APC \cite{apc}) reconstruct or predict speech signals, potentially preserving non-linguistic attributes such as speaker characteristics. Contrastive methods such as CPC \cite{cpc} and wav2vec series \cite{wav2vec,wav2vec2} distinguish positive samples from negative ones. Although effective, they often assume strict temporal correspondence. Predictive approaches (e.g., HuBERT \cite{hubert}, wavLM \cite{wavlm}, Data2vec \cite{data2vec}) rely on masked frame prediction objectives. While advancing the state-of-the-art, their frame-wise consistency reliance can cause speed perturbation sensitivity.

\subsection{Siamese Networks for Self-Supervised Learning} Siamese-based frameworks have become prominent in SSL, particularly in computer vision \cite{simclr, byol, simsiam, moco, dino}. These frameworks learn robust, invariant embeddings by comparing augmented input views via contrastive or distillation objectives.

In speech, several recent works \cite{spin, rspin, csiam, dinosr} have explored Siamese-based approaches. SPIN and R-SPIN \cite{spin, rspin}, inspired by SwaV \cite{swav}, employ iterative clustering under speaker perturbation but rely on frame-level contrast. C-Siam \cite{csiam} introduces dual encoders and a contrastive objective that assumes consistent temporal correspondence via predefined framewise alignment. DinoSR \cite{dinosr} combines masked pseudo-label prediction with online clustering and self-distillation, but also depends on frame-level alignment. While promising, their reliance on strict temporal matching can reduce robustness to varying speaking rates. LASER \cite{laser} partially addresses this limitation through soft-DTW \cite{soft-dtw} for direct sequence comparison. In contrast, our approach combines a Siamese framework with CTC-based alignment prediction, offering a more flexible alignment mechanism.

\section{SiamCTC}
\label{sec:siamctc}

This section introduces SiamCTC, a novel framework for learning speech representations invariant to temporal variations by combining a Siamese encoder with a Connectionist Temporal Classification (CTC) \cite{ctc} alignment head. Auxiliary alignment consistency and temporal contrastive losses further refine alignment and prevent collapse. The overall goal is to train an encoder that captures linguistic representations whose sequential structure remains invariant to temporal perturbations. Figure \ref{fig:siamctc} provides an overview of our approach.

\subsection{Siamese Encoder with Temporal Perturbation}

Given an input speech utterance \(X\), we create two views: the original sequence \(X\) and its temporally perturbed version \(\tilde{X}\). Both sequences are processed by a shared encoder \(f_\theta\) to obtain their respective representations:
\setlength{\abovedisplayskip}{5pt}
\setlength{\belowdisplayskip}{5pt}
\begin{equation}
    Z = f_\theta(X), \quad \tilde{Z} = f_\theta(\tilde{X}).
\end{equation}

Our approach learns flexible inter-view alignments without strict, predefined frame-level correspondence. This design enables effective use of temporal augmentations, improving robustness to natural variations in speech rate.

\subsection{Monotonic Alignment Learning} Connectionist Temporal Classification (CTC) \cite{ctc} aims to maximize the log-likelihood between two sequences with unknown alignments, a common scenario in automatic speech recognition \cite{ctc1, adamer}. Specifically, CTC sums the likelihoods of all valid alignment paths, leveraging a forward-backward procedure for efficient computation.

In our framework, the original representation $Z$ serves as the target sequence or `pseudo-label' for the CTC loss. The perturbed view representation $\tilde{Z}$ is passed through a linear prediction layer $\phi$ to produce predicted representations. Frame-wise logits for CTC are then derived from the attention scores between $Z$ and $\phi(\tilde{Z})$. These logits, forming an attention score matrix, are used in the CTC loss:
\setlength{\abovedisplayskip}{5pt}
\setlength{\belowdisplayskip}{5pt}
\begin{equation}
    \mathcal{L}_{\text{CTC}} = -\log \sum_{\pi \in \mathcal{B}^{-1}(Z)} p(\pi|\phi(\tilde{Z})),
\end{equation}
where $\mathcal{B}$ is the CTC many-to-one mapping and $\pi$ represents valid alignment path. This allows CTC to learn alignments without requiring pre-defined discrete frame-wise labels. 

For practical implementation, since CTC typically expects the input length to be longer than or equal to the target length, we \textit{transpose} our logits when necessary to fit this requirement (e.g., under speed-up perturbation). Furthermore, to handle the mandatory \verb+<blank>+ symbol in CTC decoding, we reserve a fixed \verb+<blank>+ column in our attention logits.

\subsection{Temporal Contrastive Loss} A trivial solution to the CTC objective is to collapse all frames into a single representation, thereby minimizing alignment cost without preserving fine-grained distinctions. To avoid this and ensure that distinct representations are learned for temporally distinct speech segments, we incorporate a temporal InfoNCE (TINCE) loss:
\setlength{\abovedisplayskip}{5pt}
\setlength{\belowdisplayskip}{5pt}
\begin{equation}
    \mathcal{L}_{\text{TINCE}}(h) = \mathbb{E}_{i,\mathcal{N}}\left[-\log \frac{e^{s(\tilde{h}_i, h_{i+1})/\tau}}{e^{s(\tilde{h}_i, h_{i+1})/\tau}+\sum_{j \in \mathcal{N}} e^{s(\tilde{h}_i, h_j)/\tau}}\right],
\end{equation}
where $h=[h_1, h_2, ...]$ is the input feature sequence, $\tilde{h}=\psi(h)$ is a linear prediction of anchor frame $h_i$, $\tau$ is a temperature hyperparameter,  $s(a,b) = a^\top b/\Vert a\Vert \Vert b\Vert $ is the cosine similarity, and $\mathcal{N}$ is a set of $M$ negative samples drawn from positions at least $K$ steps away from $i$. The core idea is that adjacent frames (positive pairs) should be more similar than randomly sampled non-adjacent frames (negative pairs). By pushing apart non-adjacent features, TINCE helps prevent the representation from collapsing into a single vector and preserves fine-grained adjacency between frames, which we refer to as ``local context.''

As a final loss, we compute the TINCE loss separately for both branches $Z$ and $\tilde{Z}$ and then take the average: 
\setlength{\abovedisplayskip}{5pt}
\setlength{\belowdisplayskip}{5pt}
\begin{equation}
    \mathcal{L}_\text{TINCE} = \frac{1}{2} (\mathcal{L}_\text{TINCE}(Z) + \mathcal{L}_\text{TINCE}(\tilde{Z}))
\end{equation}

\subsection{Alignment Consistency Loss} While CTC can learn monotonic alignments, its alignment quality may degrade without sufficient guidance. To address this, we utilize KL divergence loss $\mathcal{L}_{\text{KLD}}$ between the hard alignment path ($p_{\text{hard}}$), obtained via Viterbi \cite{viterbi} or DTW \cite{dtw}, and the attention scores ($p_{\text{soft}}$):
\setlength{\abovedisplayskip}{5pt}
\setlength{\belowdisplayskip}{5pt}
\begin{equation}
    \mathcal{L}_{\text{KLD}} = \mathbb{D}_\text{KL}(p_{\text{hard}} \Vert  p_{\text{soft}}).
\end{equation}

This loss constrains the learned attention to remain close to a meaningful alignment reference. In practice, we find that $\mathcal{L}_{\text{KLD}}$ has a relatively minor effect, as the TINCE loss already discourages representation collapse and implicitly guides the soft alignment away from extreme solutions.

\subsection{Overall Objective} We combine these losses into a single training objective:
\setlength{\abovedisplayskip}{5pt}
\setlength{\belowdisplayskip}{5pt}
\begin{equation}
    \mathcal{L} = \mathcal{L}_{\text{CTC}} + \alpha\mathcal{L}_{\text{KLD}} + \beta\mathcal{L}_{\text{TINCE}},
\end{equation}
where $\alpha$ and $\beta$ are balancing coefficients. We set $1.0$ for both coefficients. This composite objective simultaneously encourages robust monotonic alignment, alignment consistency, and non-collapsing temporal structure.

\section{Experimental Details}

\subsection{Datasets} We use LibriSpeech \cite{librispeech} for both pre-training and downstream fine-tuning. LibriSpeech is a widely used corpus of approximately 1,000 hours of read English audiobooks derived from LibriVox, sampled at 16 kHz. It is partitioned into subsets designated as ``clean" or ``other," reflecting differences in recording quality and speaker accents. In this work, we focus on the \textit{train-clean-100} split to facilitate rapid experimentation and training efficiency under limited computational resources. For evaluation, we use the \textit{test-clean} subset.

\subsection{Model Training}
We use two pre-trained self-supervised speech encoders as our base: HuBERT \cite{hubert} and WavLM \cite{wavlm}. Both models were originally trained on 960 hours of LibriSpeech \cite{librispeech} corpus, and are publicly available via S3PRL.\footnote{\url{https://huggingface.co/s3prl/converted_ckpts/resolve/main/hubert_base_ls960.pt}}\footnote{\url{https://huggingface.co/s3prl/converted_ckpts/resolve/main/wavlm_base.pt}} The model is then further trained for 5,000 updates, with a total batch size of 8. We employ AdamW optimizer \cite{adamw} with a peak learning rate of 2e-5, linearly warmed up over the first 1,000 updates and then linearly decayed thereafter. A maximum gradient norm of 1.0 is applied to stabilize training.

The encoder architecture is based on HuBERT \cite{hubert} and WavLM \cite{wavlm}, and outputs 256-dimensional features. The alignment head uses a temperature of $2.0$. For hard alignment, we rely on the Viterbi algorithm \cite{viterbi}. For temporal contrastive learning, we use $-1$ as \verb+<blank>+ log probability, sample 20 negative examples from positions at least 5 frames away and apply a contrastive temperature of $0.2$. For data augmentation, we apply speed perturbations using factors $\{0.8, 0.9, 1.0, 1.1, 1.2\}$ and pitch perturbations using factors $\{-2, -1, 0, 1, 2\}$.

\subsection{Baselines} We compare our approach with two widely used self-supervised speech representation encoders, HuBERT \cite{hubert} and WavLM \cite{wavlm}. In addition, we compare ours with their SPIN \cite{spin} and LASER \cite{laser} variants, which are fine-tuned on the same base models as ours. All models are evaluated under identical conditions using the SUPERB benchmark for fair comparison.\footnote{\url{https://github.com/s3prl/s3prl}}

\begin{table}[t]
\centering
\caption{Performance comparison on phoneme recognition (PR) and automatic speech recognition (ASR) downstream tasks. We report Phoneme Error Rate (PER\%) and Word Error Rate (WER\%), with lower values indicating better performance. The numbers in parentheses show the performance gap relative to the base model.}
\label{tab:sota}
\begin{tabular}{l|ll}
\toprule
\multirow{2}{*}{Model} 
& \multicolumn{1}{c}{PR}  
& \multicolumn{1}{c}{ASR} \\

& \multicolumn{1}{c}{PER(\%) $\downarrow$} 
& \multicolumn{1}{c}{WER(\%) $\downarrow$} \\
\midrule
HuBERT \cite{hubert}          & 5.41              & 6.42 \\
HuBERT+Spin \cite{spin}     & 4.39 (-1.02)             & 6.34 (-0.08) \\  
HuBERT+LASER \cite{laser}   & 4.61 (-0.80)             & 6.18 (-0.24) \\
HuBERT+SiamCTC  & 4.32 (-1.09)             & 6.23 (-0.19) \\
\midrule
WavLM \cite{wavlm}           & 4.84              & 6.21 \\
WavLM+Spin \cite{spin}      & 4.18 (-0.66)             & 5.88 (-0.33) \\
WavLM+LASER \cite{laser}     & 4.28 (-0.56)             & 5.92 (-0.29) \\
WavLM+SiamCTC   & 3.96 (-0.88)             & 5.73 (-0.48) \\
\bottomrule
\end{tabular}
\vspace{-0.3cm} 
\end{table}

\subsection{Evaluation Metrics} We report Phoneme Error Rate (PER\%) and Word Error Rate (WER\%) on the LibriSpeech \textit{test-clean} for phoneme recognition and ASR tasks, respectively. In addition, we conduct ablation studies to examine the contribution of each loss component. We also test model robustness against speaking rate variations in speed-perturbed versions of the evaluation set.

\section{Experiment Results}

\subsection{Main Results}

Table \ref{tab:sota} summarizes our results in phoneme recognition (PR) and automatic speech recognition (ASR) on LibriSpeech, comparing HuBERT \cite{hubert} and WavLM \cite{wavlm} baselines with variants enhanced by SPIN \cite{spin}, LASER \cite{laser}, or our proposed SiamCTC. We report phoneme error rate (PER\%) and word error rate (WER\%), with lower values indicating better performance. The values in parentheses show the improvement relative to the respective baseline.

For HuBERT-based systems, SiamCTC achieves the lowest PER (4.32\%), surpassing both SPIN (4.39\%) and LASER (4.61\%). Although LASER slightly outperforms SiamCTC in WER (6.18\% vs. 6.23\%), our method consistently produces a strong overall result.

In the WavLM-based setting, SiamCTC again shows the largest improvement in PER (down from 4.84\% to 3.96\%) compared to SPIN (4.18\%) and LASER (4.28\%). Notably, SiamCTC also attains the best WER (5.73\%) among the WavLM variants. These findings suggest that our explicit monotonic alignment objective provides robust benefits in capturing linguistic content, even under varying speaking rates.

\subsection{Ablation Study} Table \ref{tab:ablation} presents an ablation study examining the effect of each loss component on phoneme recognition performance with HuBERT base model. We evaluate different configurations of CTC loss, alignment consistency loss (KLD), and Temporal InfoNCE (TINCE) loss using the Phoneme Error Rate (PER\%), where lower values indicate better performance.

A baseline model employing only CTC achieves a PER of 5.26\%. Adding the KLD term to guide alignment consistency reduces PER to 5. 16\%, indicating that aligning the learned attention with a hard alignment reference provides moderate improvements. In contrast, replacing KLD with TINCE yields a more substantial gain, lowering the PER to 4.48\%. This result suggests that temporal contrastive learning effectively preserves local context and mitigates representation collapse, even without explicit hard alignment guidance.

Finally, combining CTC, KLD, and TINCE further increases performance to 4.32\%, demonstrating that alignment consistency and temporal contrastive learning complement each other. In general, these findings confirm that both KLD and TINCE play a significant role in improving the robustness and accuracy of the learned speech representations.

\begin{table}[t]
\centering
\caption{Ablation study on different loss components for phoneme recognition. We evaluate performance in terms of Phoneme Error Rate (PER\%) using combinations of CTC, KL divergence (KLD), and Temporal InfoNCE (TINCE) loss. Lower PER indicates better performance.
}
\label{tab:ablation}
\begin{tabular}{ccc|c}
\toprule
\multicolumn{3}{c|}{Loss} & PR \\
CTC & KLD & TINCE & PER(\%) $\downarrow$ \\
\midrule
\cmark & \xmark & \xmark & 5.26 \\
\cmark & \cmark & \xmark & 5.16 \\
\cmark & \xmark & \cmark & 4.48 \\
\cmark & \cmark & \cmark & 4.32 \\
\bottomrule
\end{tabular}
\vspace{-0.3cm} 
\end{table}

\subsection{Analysis}

\begin{figure}[t!]
    \centering
    \includegraphics[width=\linewidth]{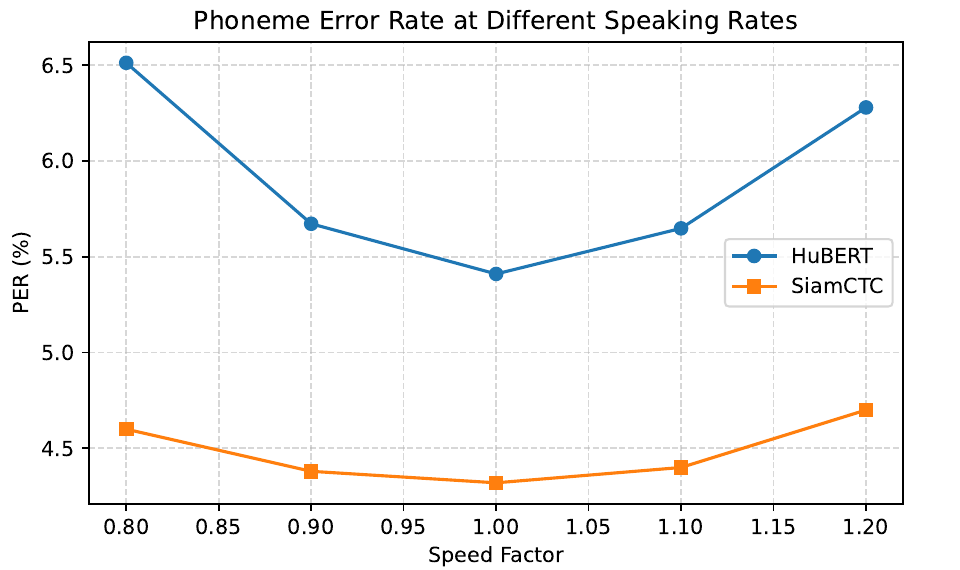}
    \caption{Phoneme Error Rate (PER) on LibriSpeech \textit{test-clean} for different speed factors. The blue circle line represents HuBERT, and the orange square line represents SiamCTC.}
    \label{fig:speed_plot}
    \vspace{-0.5cm}
\end{figure}

We evaluate the fine-tuned HuBERT and its SiamCTC variant on speed factors $\{0.8, 0.9, 1.0, 1.1, 1.2\}$ to assess the robustness of our framework against different speaking rates. Figure~\ref{fig:speed_plot} shows the Phoneme Error Rate (PER) for each factor, where lower values indicate better performance.

Across all speed factors, SiamCTC consistently achieves a lower PER than HuBERT. At the original speed (1.0), HuBERT obtains a PER of 5.41\%, whereas SiamCTC achieves 4.32\%. When the audio is slowed down, HuBERT’s PER rises above 5.60\%, peaking at 6.52\% at $0.8\times$. In contrast, SiamCTC exhibits a smaller performance drop, staying below $4.6\%$. For faster speech, HuBERT performance again degrades to $6.28\%$ at $1.2\times$, whereas SiamCTC remains relatively stable (4.7\% at $1.2\times$). These results suggest that learning temporal perturbation invariance produces more robust representations across various speaking speeds and styles.

\section{Conclusion}

We proposed SiamCTC, a self-supervised learning framework that merges Siamese encoding with a Connectionist Temporal Classification (CTC) based alignment objective as its core mechanism to handle temporal perturbations in speech. This is complemented by temporal InfoNCE (TINCE) loss and an alignment consistency loss to prevent representation collapse and refine alignment quality. SiamCTC provides flexible alignment without relying on strict frame-level matching, thus capturing content invariance across varying speaking rates. Experiments on the LibriSpeech dataset show that SiamCTC demonstrates notable performance gains over established baselines, suggesting its effectiveness in preserving linguistic content even under speed perturbations. Our findings emphasize the benefits of learning flexible monotonic alignments in a self-supervised manner, paving the way for more robust and adaptive speech representation learning.

\section{Limitation}
While SiamCTC has shown promising performance, we observe that the downstream results can be sensitive to hyperparameters such as augmentation strategies, negative pair sampling, and attention logit temperature. In particular, we find that using a lower temperature, which produces more peaked logits, is critical for alignment seeking. Future work would address these sensitivities with more robust or adaptive strategies.

While current SiamCTC is built on pre-trained models for efficiency, training SiamCTC from scratch may further reveal its full potential--especially if combined with broader augmentation techniques like masking. Exploring discrete speech units (e.g., via Vector Quantization \cite{vqvae}) for applications like spoken language modeling \cite{discrete-slm}, beyond current continuous representations, can be another future direction.

\section{Acknowledgements}
This work was partly supported by Center for Applied Research in Artificial Intelligence (CARAI) grant funded by DAPA and ADD (UD230017TD) and partly supported by Institute of Information \& communications Technology Planning \& Evaluation (IITP) grant funded by the Korea government(MSIT) (No.RS-2022-II220184, Development and Study of AI Technologies to Inexpensively Conform to Evolving Policy on Ethics).

\bibliographystyle{IEEEtran}
\bibliography{mybib}

\end{document}